\documentclass[10pt]{article}
\usepackage{graphicx}  
\usepackage{dcolumn}   
\usepackage{bm}        
\usepackage{verbatim}  

\newtheorem{myProperty}{Property}

\newcommand{\state}[1]{\left|{#1}\right>}
\newcommand{\stateBasis}[2]{\left|{#1}\right>_{#2}}

\begin{document}

\title{Is independent choice possible?}
\author{Hitoshi Inamori\\
\small\it Soci\'et\'e G\'en\'erale\\
\small\it Boulevard Franck Kupka, 92800 Puteaux, France\\
\small\it Previous academic affiliation: \\
\small\it Centre for Quantum Computation, Clarendon Laboratory, Oxford University
}


\bigskip

\date{\today}

\maketitle

\begin{abstract}
This paper questions the generally accepted assumption that one can make a random choice that is independent of the rest of the universe. We give a general description of any setup that could be conceived to generate random numbers. Based on the fact that the initial state of such setup together with its environment cannot be known, we show that the independence of its generated output cannot be guaranteed. Some consequences of this theoretical limitation are discussed.
\bigskip


\end{abstract}

\section{Introduction}
We are ultimately able to make choices, independently of the world that surrounds us.
This is {\it free will}, a belief which is so deeply rooted in our culture, that we generally take it for granted without even noticing.  
The assumption that one can make an independent decision is however crucial in many logical reasonings and algorithms. 

Making a decision is a physical process. Being able to make a random choice implies that there is a physical setup that provides random outcomes, which can be represented -- without loss of generality -- as numbers. Independent random choice implies the existence of a physical setup that provides random numbers that are independent of the rest of the universe.  

It is already known that one can {\it expand} an initial random number thanks to quantum algorithms known as Private Randomness Expansion~\cite{Colbeck, ColbeckKent, Pironio,Vazirani}, whose aim is to increase (in entropy) randomness using untrusted devices. 
Private Randomness Expansion does however not deal with the generation of the initial random number, as it is proved~\cite{Colbeck} that one cannot generate random number from scratch with untrusted devices, but can only expand an initial random number.

The scope of this paper is different: we study the feasibility of actual {\it generation} of random numbers using trusted devices, which does not assume the presence of initial random numbers serving as a seed.
Based on the law of quantum mechanics, and the fact that the state of the overall system including the setup and its environment can never be known, we prove that such generation cannot be guaranteed to be truly independent of the rest of the universe.

The paper is organized as follows: we start by considering a simple setup that is usually believed to generate independent random numbers. We show however that such setup can be entangled with its environment prior to the experiment, and as such, the random numbers returned by the setup can never be guaranteed in theory to be independent of the environment. 

We show that this result can be generalized to any physical setup that could be conceived to generate random numbers. Independence of random number generation cannot be guaranteed, and we discuss the theoretical consequences of such limitation.

\section{Case study}
Throughout this paper, our question will be the following: can we conceive a source of random binary number that is guaranteed to be independent of the rest of the universe? 

The usual setup that comes to mind when one wants a random source of binary number is the following. Consider a qubit and let $\{\state{0},\, \state{1}\}$ be its canonical basis. If we had a certified way of preparing the qubit in the state $\state{+}$ where $\state{\pm}=\frac{1}{\sqrt{2}}\left(\state{0}\pm\state{1}\right)$, then we could measure such qubit in the basis $\{\state{0},\, \state{1}\}$. If the state is observed in the state $\state{0}$ then we say that the setup generated the random number 0. If the state is observed in the state $\state{1}$ then the random number is 1. The process is guaranteed to generate a random bit that is independent of the rest of the universe.

But how do we prepare the state $\state{+}$ in practice? The setup that is usually employed works as follows:
The measured qubit is prepared in some state, and is first observed in the $\{\stateBasis{+}{M},\, \stateBasis{-}{M}\}$ basis. 
We know that the measured qubit is now in either the $\stateBasis{+}{M}$ state or $\stateBasis{-}{M}$ state, 
information which must be explicitly kept (otherwise the effect of the measurement can be cancelled, as shown in the Quantum Eraser experiment~\cite{QuantumEraser}). 
This state is then measured in the $\{\stateBasis{0}{M},\, \stateBasis{1}{M}\}$ basis.

The whole setup can be represented as in Figure~\ref{Fig1}:
\setlength{\unitlength}{0.025 in}
\begin{figure}[!h]
\centering

\begin{picture}(100,50)

\put(15,40){\line(1,0){10}}
\put(35,40){\line(1,0){20}}
\put(65,40){\line(1,0){10}}

\put(25,35){\framebox(10,10){$H$}}
\put(45,40){\circle*{4}}
\put(55,35){\framebox(10,10){$H$}}

\put(45,40){\line(0,-1){28}}

\put(45,15){\circle{6}}

\put(15,15){\line(1,0){60}}

\put(10,40){\makebox(0,0){$M$}}

\put(10,15){\makebox(0,0){$P$}}

\put(95,40){\makebox(0,0){\small measurement}}

\put(95,15){\makebox(0,0){\small measurement}}

\end{picture}
\caption{An usual setup for random bit generation}\label{Fig1}
\end{figure}
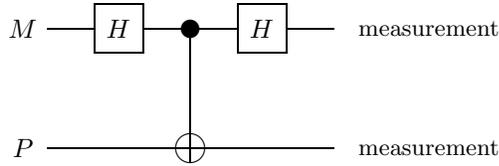
The qubit $P$ represents the outcome of the first measurement ($P$ stands for ``projection''), 
whereby the first qubit is measured in the $\stateBasis{+}{M}$ state or the $\stateBasis{-}{M}$ state. 
We look at the value returned from the measurement of $M$, given the outcome of the measurement of $P$.

Note that although we find it natural to describe the measurement of $P$ as being prior to the measurement of $M$, it actually does not matter whether $M$ is measured before or after $P$.

With such a setup, we generally assume that measurement of $M$ leads to a random binary outcome, that is completely independent of the rest of the universe. 
This belief is natural if one accepts that the initial state of the setup $M\otimes P$ is known, or at least, that the initial state of the setup is not entangled with the rest of the universe.

However, in theory, the initial quantum state of a given experimental setup is never known. 
One can choose arbitrarily the basis for the Hilbert space describing the system under study, 
but one cannot guarantee that the system under study is not entangled with the rest of the universe. 
Said differently, there is no theoretical guarantee that a physical system under study had not been entangled at an earlier point with another part of the universe.

Suppose for instance that the setup above was entangled initially with a third qubit, that is part of the rest of the universe and is not known to us. 
Let's denote this third qubit $E$ (Figure~\ref{Fig2}).

\setlength{\unitlength}{0.025 in}
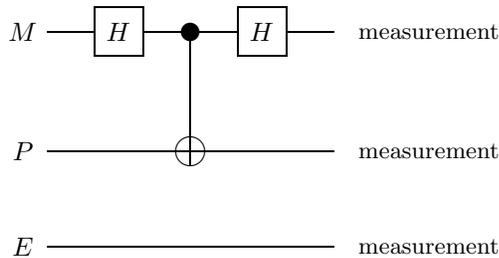
\begin{figure}[!h]
\centering
\begin{picture}(100,70)

\put(15,60){\line(1,0){10}}
\put(35,60){\line(1,0){20}}
\put(65,60){\line(1,0){10}}

\put(25,55){\framebox(10,10){$H$}}
\put(45,60){\circle*{4}}
\put(55,55){\framebox(10,10){$H$}}

\put(45,60){\line(0,-1){28}}

\put(45,35){\circle{6}}

\put(15,35){\line(1,0){60}}

\put(15,15){\line(1,0){60}}

\put(10,60){\makebox(0,0){$M$}}

\put(10,35){\makebox(0,0){$P$}}

\put(10,15){\makebox(0,0){$E$}}

\put(95,60){\makebox(0,0){\small measurement}}

\put(95,35){\makebox(0,0){\small measurement}}

\put(95,15){\makebox(0,0){\small measurement}}

\end{picture}
\caption{The same setup taking into account the environment}\label{Fig2}
\end{figure}

Suppose now that the initial state of the resulting system is:

\begin{eqnarray}
\state{\alpha_0} = \frac{1}{\sqrt{4}}&\big(&
	\stateBasis{0}{M}\otimes\stateBasis{0}{P}\otimes\stateBasis{0}{E} \nonumber\\ 
	&+&\stateBasis{0}{M}\otimes\stateBasis{1}{P}\otimes\stateBasis{0}{E} \nonumber\\
	&+&\stateBasis{1}{M}\otimes\stateBasis{0}{P}\otimes\stateBasis{1}{E} \nonumber\\
	&+&\stateBasis{1}{M}\otimes\stateBasis{1}{P}\otimes\stateBasis{1}{E}\big),
\end{eqnarray}
then, calculation shows that the final state of the system is, after going through the experimental setup:

\begin{eqnarray}
\state{\alpha} = \frac{1}{\sqrt{4}}&\big(&
	\stateBasis{0}{M}\otimes\stateBasis{0}{P}\otimes\stateBasis{0}{E} \nonumber\\ 
	&+&\stateBasis{1}{M}\otimes\stateBasis{0}{P}\otimes\stateBasis{1}{E} \nonumber\\
	&+&\stateBasis{0}{M}\otimes\stateBasis{1}{P}\otimes\stateBasis{0}{E} \nonumber\\
	&+&\stateBasis{1}{M}\otimes\stateBasis{1}{P}\otimes\stateBasis{1}{E}\big),
\end{eqnarray}
which is identical to the initial state and which can be rewritten as:

\begin{eqnarray}
\state{\alpha} &=& \frac{1}{\sqrt{4}}\big(
	\stateBasis{0}{M}\otimes\stateBasis{0}{E}+\stateBasis{1}{M}\otimes\stateBasis{1}{E}\big)\otimes\stateBasis{0}{P}\nonumber\\
&+&\frac{1}{\sqrt{4}}\big(
	\stateBasis{0}{M}\otimes\stateBasis{0}{E}+\stateBasis{1}{M}\otimes\stateBasis{1}{E}\big)\otimes\stateBasis{1}{P}.
\end{eqnarray}

In other words, whichever result is observed for the projection qubit $P$, the measured qubit $M$ and the environment qubit $E$ are perfectly entangled. Measurement outcome at $M$ and $E$ will be perfectly correlated if we use the same measurement basis for $M$ and $E$. 

We see that the proposed setup $M\otimes P$ which is usually used as a random and independent source of random number, cannot actually be guaranteed to be independent of the rest of the universe.

\section{General case}
We saw in the section above that a setup that was believed to generate independent random choices could actually not be guaranteed to be independent. 

Our next question is: can any setup in general produce random numbers that are guaranteed to be independent of the rest of the universe?

However complex such a setup can be, it can be described as follows: the setup is a physical system, made of two subcomponents $M$ and $P$, that interact and evolve possibly in a most general and complex manner. The subcomponents $M$ and $P$ are then observed.
 We denote by $U$ the unitary map describing the evolution of $M$ and $P$. The outcome of these observations are denoted $m$ and $p$ respectively. The generation of the random number is deemed successful if $p$ is part of a pre-agreed set $\mathcal{S}$ corresponding to \lq\lq{}successful\rq\rq{} outcomes. In such a case, the random number is given by a pre-agreed deterministic function of $m$ and $p$, $f(m,p)$. The setup is represented in Figure~\ref{Fig3}.

\setlength{\unitlength}{0.025 in}
\begin{figure}[!h]
\centering
\begin{picture}(150,70)

\put(15,13){\line(1,0){10}}
\put(17,10){\makebox(5,20){$\vdots$}}
\put(15,25){\line(1,0){10}}
\put(15,43){\line(1,0){10}}
\put(17,40){\makebox(5,20){$\vdots$}}
\put(15,55){\line(1,0){10}}

\put(25,10){\framebox(20,50){$U$}}

\put(45,13){\line(1,0){10}}
\put(47,10){\makebox(5,20){$\vdots$}}
\put(45,25){\line(1,0){10}}
\put(45,43){\line(1,0){10}}
\put(47,40){\makebox(5,20){$\vdots$}}
\put(45,55){\line(1,0){10}}

\put(5,50){\makebox(0,0){$M$}}

\put(5,20){\makebox(0,0){$P$}}

\put(80,50){\makebox(0,0){\small measurement $\mapsto m$}}

\put(80,20){\makebox(0,0){\small measurement $\mapsto p$}}

\put(105,50){\line(1,0){20}}
\put(105,20){\line(1,0){20}}
\put(125,50){\line(0,-1){30}}
\put(125,35){\line(1,0){10}}

\put(145,37){\makebox(0,0){\small $f(m,p)$}}
\put(145,30){\makebox(0,0){\small if $p\in\mathcal{S}$}}

\end{picture}
\caption{General setup for random number generation}\label{Fig3}
\end{figure}
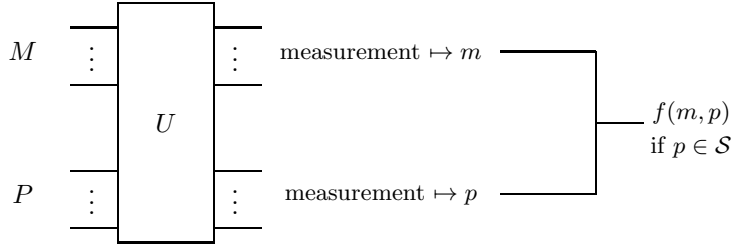

Now, whatever the size of the setup, we assume that it is sufficiently small compared to the entire universe. Consequently we can define a quantum system $E$ that is part of the rest of the universe, described by a Hilbert space of the same dimensionality as $M$. We denote by $V=U\otimes I_E$ the unitary operator that applies $U$ on the system $M\otimes P$ and leaves $E$ unchanged (Figure~\ref{Fig4}).

Consider the following state for the combined physical system made of $M$, $P$ and $E$:

\begin{equation}
\state{\Psi} = \mathcal{N} \sum_{k} \stateBasis{k}{M}\otimes \stateBasis{\phi_0}{P} \otimes \stateBasis{k}{E}.
\end{equation}
where $\stateBasis{\phi_0}{P}$ is a state for $P$ that leads to a successful fixed outcome $p\in\mathcal{S}$, and $\{\stateBasis{k}{M}\}$ and $\{\stateBasis{k}{E}\}$ are bases for $M$ and $E$, respectively. The number $\mathcal{N}$ is a normalisation factor.

By construction, if the initial state of the entire system $M\otimes P\otimes E$ is in the state $\state{\Psi_0}=V^{-1}\state{\Psi}$, then the experiment returns the state $\state{\Psi}$ in which $M$ and $E$ are perfectly entangled, and with which the measurement at $P$ leads to a successful result $p\in\mathcal{S}$. Measurement of $E$ leads to the same outcome as the measurement of $M$, and the outcome $f(m,p)$ can be completely deduced from the measurement outcome of $E$. 
Therefore, there exists at least one initial state $\state{\Psi_0}$ with which the proposed setup does not return independent random numbers.

\setlength{\unitlength}{0.025 in}
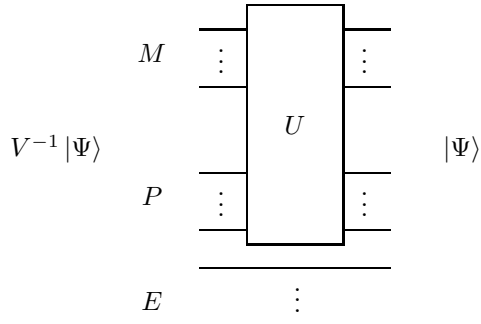
\begin{figure}[!h]
\centering
\begin{picture}(100,90)

\put(25,33){\line(1,0){10}}
\put(27,30){\makebox(5,20){$\vdots$}}
\put(25,45){\line(1,0){10}}
\put(25,63){\line(1,0){10}}
\put(27,60){\makebox(5,20){$\vdots$}}
\put(25,75){\line(1,0){10}}

\put(35,30){\framebox(20,50){$U$}}

\put(55,33){\line(1,0){10}}
\put(57,30){\makebox(5,20){$\vdots$}}
\put(55,45){\line(1,0){10}}
\put(55,63){\line(1,0){10}}
\put(57,60){\makebox(5,20){$\vdots$}}
\put(55,75){\line(1,0){10}}

\put(25,13){\line(1,0){40}}
\put(43,10){\makebox(5,20){$\vdots$}}
\put(25,25){\line(1,0){40}}

\put(15,70){\makebox(0,0){$M$}}

\put(15,40){\makebox(0,0){$P$}}

\put(15,18){\makebox(0,0){$E$}}

\put(80,50){\makebox(0,0){$\state{\Psi}$}}

\put(-5,50){\makebox(0,0){$V^{-1}\state{\Psi}$}}

\end{picture}
\caption{General setup taking into account environment}\label{Fig4}
\end{figure}

As such no physical setup can guarantee generation of independent random numbers. We have demonstrated the following:

\begin{myProperty} No generation of random choices can be guaranteed to be independent of the remaining part of the universe.
\end{myProperty}

\section{Consequences}
We have proven that -- at least in theory -- no random choice can be guaranteed to be independent of the rest of the universe. This is mainly due to the fact that, if we consider the state of any physical setup together with its environment, then such state is unknown. One could argue that, statistically, no relevant impact can be expected from these entanglements that are only theoretical possibilities. One could also argue that the likelyhood that the overall state of the universe is such that it precisely introduces dependence between the generated random number and its environment is dim.
However, we cannot rely on statistical argument when independence of random choices is itself under question.

The fact that we cannot guarantee independence of random choices has many consequences, few of them are discussed below. 

\subsection{On the Positive-Operator Valued Measure} 
Positive-Operator Valued Measure~\cite{POVM} is considered to be the most general formulation of a measurement in quantum mechanics. 
It allows the introduction of probability weights between different projective measurements, as if it was possible to choose randomly between these projective measurements. However, we have just seen that the random choice between these different projective measurements cannot be guaranteed to be independent of the rest of the universe. Therefore, the mathematical description using POVM is misleading, in the sense that it introduces the false belief that independent random choice is possible. Our view in this paper is that the most fundamental description of measurement remains the law based on projection of quantum states onto the measurement basis.

\subsection{On Bell's theorem and its interpretation} 
The experimental violation of Bell's inequality~\cite{Aspect} is generally considered as a confirmation that laws of nature cannot be described by local hidden variables. However, existence of local hidden variable is not the only assumption which is necessary in the derivation of Bell’s inequality~\cite{Bell, Bell71}. In particular, Bell assumes that random and independent decisions can be taken at the two distant locations. We suggest in this paper that, what could be forbidden by the violation of the Bell's inequality is actually not the existence of local hidden variable, but rather the true independence between the observed system and the way that system is observed. In other words, violation of Bell's inequality could be interpreted as the demonstration that one cannot make decisions that are fully independent of the environment.

\subsection{On algorithms relying on the independence of random choices} 
Many algorithms, such as the Monte Carlo integration, error-correction and cryptography to name a few, rely on the ability to generate random numbers that are independent. The theoretical possibility that some remaining part of the universe may be correlated to these random numbers may, or may not be relevant. In any case, for theoretical completeness, one should not take for granted that sources of random choices are truly independent of the rest of the universe. Impact of potential dependence of such sources with the rest of the universe should be analyzed.     

\section{Discussion}
We generally assume that one can take a random decision (using if necessary a device as described in the case study above), and that such decision can be taken independently of all the rest of the universe. This is {\it free-will}, this commonly shared belief that if one wants, one is able to decide by oneself, independently of its environment. 

We have shown in this paper that no random choice can be guaranteed to be independent of the rest of the universe. For practical purposes, the theoretical possibility that one part of the universe may be correlated with our experimental setup may have little importance. This said, if we believe that our current universe expanded from a single common state, then we cannot dismiss the possibility that any experimental setup under study is entangled with another observable part of the universe.

\section{Acknowledgements}
The author thanks Norbert L\"utkenhaus for his comments on the initial version of this paper and for pointing to the existing work on Private Randomness Expansion.

\end{document}